\documentclass[aps,prl,twocolumn,showpacs,preprintnumbers,amsmath,amssymb,floatfix]{revtex4}

\usepackage{graphicx}
\usepackage{dcolumn}
\usepackage{bm}

\newcommand {\beq} {\begin{equation}}
\newcommand {\eeq} {\end{equation}}
\newcommand {\beqa}{\begin{eqnarray}}
\newcommand {\eeqa}{\end{eqnarray}}
\newcommand {\n} {\nonumber}

\newcommand {\tr}{{\rm tr\,}}

\begin{document}


\title{Expanding universe as a classical solution in the Lorentzian matrix model\\
for nonperturbative superstring theory}

\author{Sang-Woo Kim$^{1}$}
\email{sang@het.phys.sci.osaka-u.ac.jp}
\author{Jun Nishimura$^{2,3}$}
\email{jnishi@post.kek.jp}
\author{Asato Tsuchiya$^{4}$}
\email{satsuch@ipc.shizuoka.ac.jp}

\affiliation{
$^{1}$Department of Physics, Osaka University,
Toyonaka, Osaka 560-0043, Japan\\
$^{2}$KEK Theory Center, High Energy Accelerator Research Organization,
		Tsukuba 305-0801, Japan \\
$^{3}$Department of Particle and Nuclear Physics,
School of High Energy Accelerator Science,
Graduate University for Advanced Studies (SOKENDAI),
Tsukuba 305-0801, Japan \\
$^{4}$Department of Physics, Shizuoka University,
836 Ohya, Suruga-ku, Shizuoka 422-8529, Japan
}

\date{October 2011; preprint: KEK-TH-1503, OU-HET-729-2011}



\begin{abstract}
Recently we have shown by Monte Carlo simulation
that expanding (3+1)-dimensional universe
appears dynamically from
a Lorentzian matrix model for
type IIB superstring theory in (9+1)-dimensions.
The mechanism for the spontaneous breaking of rotational symmetry
relies crucially on the noncommutative nature of the space.
Here we study the classical equations of motion as a complementary approach.
In particular, we find a unique class of SO(3) symmetric solutions,
which exhibits the time-dependence compatible with the expanding universe.
The space-space noncommutativity is exactly zero,
whereas
the space-time noncommutativity becomes significant only towards
the end of the expansion.
We interpret the Monte Carlo results and the classical solution
as describing the behavior of the model at earlier time and at later time,
respectively.
%
\end{abstract}


\maketitle
\paragraph*{Introduction.---}

It is widely believed that the birth of our universe can be
described by superstring theory, which is a natural candidate
for a unified theory including quantum gravity.
Indeed, a lot of insights into this issue have been obtained
by string cosmology over the last decade \cite{string cosmology}.
These studies are based on perturbative formulations incorporating
nonperturbative effects through D-branes.
An obvious drawback in such an approach, however,
is that one has to choose
a particular string vacuum from numerous vacua that are theoretically allowed.
On the other hand, there is also a possibility that
one can actually determine the true string vacuum uniquely
if one uses a nonperturbative formulation.

Along this line of thought, we have
studied a SO(9,1) symmetric Lorentzian matrix model,
which is considered
to be a nonperturbative definition
of type IIB superstring theory in (9+1) dimensions \cite{Kim:2011cr}.
Surprisingly
our Monte Carlo results provide clear evidence
that 3 out of 9 directions start to expand at some critical time.
%
The observed spontaneous breaking of the SO(9) rotational symmetry
down to SO(3) has been understood intuitively by a mechanism, which
relies crucially
on the noncommutative nature of the space.
While this is certainly intriguing, it also poses a crucial
question
whether the space-time becomes commutative at later time
as we observe it now.

%

%

In this Letter we study the classical equations of motion of the model
as a complementary approach.
%
In particular, we find a unique class of SO(3) symmetric solutions,
which turns out to have
the time dependence compatible with the expanding universe.
For this solution,
the space-space noncommutativity is exactly zero,
whereas
the space-time noncommutativity becomes significant only towards
the end of the expansion.
\paragraph*{Lorentzian matrix model.---}

The matrix model proposed as a nonperturbative formulation
of type IIB superstring theory
has the action $S = S_{\rm b} + S_{\rm f}$, where \cite{IKKT}
%
\beqa
S_{\rm b} &=& -\frac{1}{4g^2} \, \tr \Bigl( [A_{\mu},A_{\nu}]
[A^{\mu},A^{\nu}] \Bigr) \ , \n  \\
S_{\rm f}  &=& - \frac{1}{2g^2} \,
\tr \Bigl( \Psi _\alpha (\, {\cal C} \,  \Gamma^{\mu})_{\alpha\beta}
[A_{\mu},\Psi _\beta] \Bigr)  \ ,
\label{action}
\eeqa
with  $A_\mu$ ($\mu = 0,\cdots, 9$) and
$\Psi_\alpha$ ($\alpha = 1,\cdots , 16$) being
$N \times N$ traceless Hermitian matrices.
The Lorentz indices $\mu$ and $\nu$ are
contracted
using the metric
$\eta={\rm diag}(-1 , 1 , \cdots , 1)$.
The $16 \times 16$ matrices $\Gamma ^\mu$ are
ten-dimensional gamma matrices after the Weyl projection,
and the unitary matrix ${\cal C}$ is the
charge conjugation matrix.
The action has manifest SO(9,1) symmetry,
where $A_{\mu}$ and $\Psi _\alpha$ transform as a
vector and
a Majorana-Weyl spinor, respectively.
The space-time is represented dynamically
by the ten bosonic matrices $A_\mu$ \cite{AIKKT}.

An important feature of the Lorentzian model is that
the bosonic part of the action is proportional to
\beq
\tr (F_{\mu\nu}F^{\mu\nu})=
- 2 \, \tr (F_{0i})^2 + \tr (F_{ij})^2 \ ,
\label{bosonic-action}
\eeq
where $F_{\mu\nu} = - i [A_\mu , A_\nu]$ are Hermitian matrices,
and hence the two terms in (\ref{bosonic-action}) have opposite signs.
A common approach to study the nonperturbative dynamics of this model
was to make the Wick rotation $A_0=i A_{10}$
and to study the SO(10) symmetric Euclidean model,
which is proved to have finite partition function \cite{Krauth:1998xh,AW}.
See Ref.~\cite{Nishimura:2011xy} and references therein
for studies of the spontaneous symmetry breaking (SSB)
of the SO(10) in the Euclidean model \cite{endnote-euclid}.
On the other hand,
it is suggested that
the Lorentzian signature of the metric
plays an important role in the dynamics of
quantum gravity \cite{Ambjorn:2005qt,Kawai:2011qb}.

In Ref.~\cite{Kim:2011cr} we studied, for the first time,
the nonperturbative dynamics of
the Lorentzian model defined by
\beq
Z = \int d A \, d\Psi \, e^{i S} =
\int d A \,  e^{i S_{\rm b}} {\rm Pf}{\cal M}(A)
\ ,
\label{partition-fn-def}
\eeq
where the Pfaffian ${\rm Pf}{\cal M}(A)$ appears from integrating
out the fermionic matrices $\Psi_\alpha$.
We
made the partition function (\ref{partition-fn-def}) finite
by introducing infrared cutoffs in both the spatial and temporal directions
instead of making the Wick rotation.
It was shown by Monte Carlo simulation
that one can remove these cutoffs in the large-$N$ limit,
and that the theory thus obtained has no parameters other than
one scale parameter.


\paragraph*{The classical solutions.---}

Taking account of the infrared cutoffs introduced in the Lorentzian model,
we search for stationary points of
the bosonic action $S_{\rm b}$ for fixed
$\frac{1}{N} \tr (A_0)^2$  and $\frac{1}{N} \tr (A_i)^2$.
Then the problem reduces to solving
the classical equations of motion
\beqa
-  [A_0,[A_0, A_i]]
+  [A_j,[A_j, A_i]] - \lambda A_i &=& 0 \ , \nonumber \\
-  [A_j,[A_j, A_0]]  - \tilde{\lambda} A_0 &=& 0 \ ,
\label{cl-eq}
\eeqa
where $\lambda$ and $\tilde{\lambda}$ represent the Lagrange multipliers
corresponding to the constraints.
We look for solutions, which are given by a unitary representation of a Lie algebra
$[A_\mu , A_\nu] = i f_{\mu\nu\lambda} A_{\lambda}$,
which guarantees automatically that the Jacobi identity is satisfied.
(See Ref.~\cite{Chatzistavrakidis:2011su} for an analogous study
in the Euclidean model.)
Motivated by the Monte Carlo results mentioned above,
we restrict ourselves to solutions
with $A_I=0$ ($4\le I \le 9$) and with SO(3) symmetry
corresponding to rotations in the $i=1,2,3$ directions.
From the complete list of real Lie algebras with
four generators $e_\mu$ ($0 \le \mu \le 3$),
the one with SO(3) symmetry is given \emph{uniquely} by
$[e_0 , e_i] = - i \, e_i$ for $i=1,2,3$, whereas
all the other commutators vanish.
(This corresponds to the algebra $A_{4,5}^{ab}$ in Table I
of Ref.~\cite{patera} for $a=b=1$.)

The unitary irreducible representations of the above algebra
are classified into two categories.
One consists of the trivial one-dimensional representations
given by $e_0=a$ and $e_i=0$, where $a$ is a real parameter.
The other consists of the infinite-dimensional representations
given by the operators
$e_0 = - i  \frac{d}{dx}$ and $e_i = a_{i}\exp (x)$
on the space of functions of $x$ with $L^2$ integrability,
where the three real parameters $a_i$ specify a representation.
As the basis of the functional space, we use the eigenfunctions
of the Hamiltonian of a one-dimensional harmonic oscillator,
which are given as
\begin{eqnarray*}
f_n(x) = c_n H_n(x) \, e^{-\frac{1}{2} x^2} \ , \quad
c_n = (\pi ^{1/4} \sqrt{n !} \, 2 ^{n/2})^{-1} \ .
\end{eqnarray*}
The representation matrices of $e_0$ and $e_i/a_i$, which we denote
as $\hat{P}$ and $\hat{K}$, respectively, have the following elements.
\begin{eqnarray*}
P_{nm} &=& \int dx f_n(x)^{*} (-i)\frac{d}{dx} f_m(x) \\
&=& - i \frac{1}{\sqrt{2}}
( \sqrt{m} \delta_{n, m-1} - \sqrt{m+1} \delta_{n, m+1}) \ , \\
\end{eqnarray*}
\begin{eqnarray*}
K_{nm} &=& \int dx f_n(x)^{*} e^{x} f_m(x)  \\
&=& c_n c_m e^{1/4} \int dx
e^{-x^2} H_n(x+\frac{1}{2}) \,
H_m (x+\frac{1}{2})  \\
&=&
e^{1/4} \, 2^{-|n-m|/2} \sqrt{n ! m!} \nonumber \\
&~& \times  \sum _{l=0}^M
\Bigl[ 2^l \,  l !  \, (M-l) ! \, (|n-m|+l) ! \Bigr]^{-1} \ ,
\end{eqnarray*}
where $M= {\rm min}(n,m)$.
In the last equality, we have used
the property $\frac{d}{dx} H_n(x) = 2n H_{n-1}$
of the Hermite polynomials.

Using a direct sum of the non-trivial representations, we find
a set of SO(3) symmetric solutions to (\ref{cl-eq}),
which is given by
\beqa
\label{A0-sol}
A_0 &=& \sqrt{\lambda} \,  \hat{P} \otimes {\bf 1}_k \ , \\
A_i &=& \hat{K} \otimes {\rm diag}(x_{1i} , \cdots , x_{ki}) \  .
\label{Ai-sol}
\eeqa
The parameters $x_{ai}\equiv ({\bf x}_{a})_i$ should be chosen such that
the points ${\bf x}_a$ ($a=1, \cdots , k$)
have spherically symmetric distribution
in the 3-dimensional space.
One of the Lagrange multiplier is fixed as $\tilde{\lambda}=0$.

In the following analysis, the $k \times k$ matrices that appear
in (\ref{A0-sol}) and (\ref{Ai-sol}) are omitted since they
only give an irrelevant constant factor.
Also we consider only one spatial direction $i=1$ for simplicity
since it turns out that the number of spatial directions does not
play any role.

\paragraph*{The space-time structure.---}

In order to extract the space-time structure from the solution,
we first need to diagonalize $A_0$.
We do this numerically by truncating the functional space
to the $N$-dimensional space spanned by $f_n(x)$ with $0 \le n \le N-1$.
Let us define the eigenvectors $| t_a \rangle$ corresponding
to the eigenvalues $t_a$ of $A_0$ ($a=1 , \cdots , N$)
with the specific order $t_1 < \cdots < t_N$.
The spatial matrix $\langle t_{I} | A_1 | t_{J} \rangle$
in that basis is not diagonal.
However, it turns out that the off-diagonal elements decay
exponentially in the direction orthogonal to the diagonal line.

To see it explicitly, let us consider the $N\times N $ matrix
$Q_{IJ} = \langle t_{I} | (A_1)^2 | t_{J} \rangle$.
Plotting $\sqrt{|Q_{IJ}/Q_{N/2,N/2}|}$
against $I-J$ for a fixed value of $(I+J)/2$,
we find that it decreases exponentially with $|I-J|$.
The half width is largest for $(I+J)/2=N/2$, and we denote it as $n$
for later convenience. For $N=16,32,64,128$, we obtain $n=11,15,23,33$.

\begin{figure}[htb]
\begin{center}
\includegraphics[height=6cm]{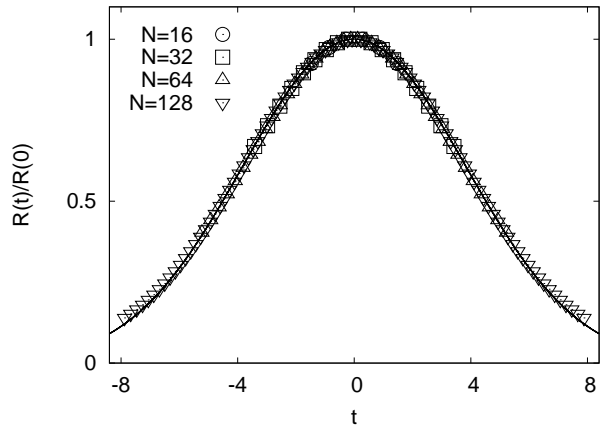}
\end{center}
\caption{
The extent of space $R(t)/R(0)$
is plotted as a function of $t$
for four values of $N$.
The block size $n$ is determined from the
the decay rate of the off-diagonal elements of $A_1$
in the basis which diagonalizes $A_0$.
The value of $\lambda$ is chosen for each $N$
in such a way that the results scale in $N$.
The solid line represents $y=\exp(-0.034 \, t^2)$,
which is obtained by fitting the $N=128$ data
to the Gaussian function.
}
\label{fig:Rt}
\end{figure}

The above observation motivates us to define $n\times n$ matrices
$\bar{A}_i^{(ab)}(t) \equiv  \langle t_{\nu+a} | A_i | t_{\nu+b} \rangle $
with $1 \le a , b \le n$ and
$t= \frac{1}{n}\sum_{a=1}^{n} t_{\nu + a}$
for $\nu=0,\cdots , (N-n)$.
These matrices represent the space structure at fixed time $t$.
Let us define the extent of space at the time $t$ as
$R(t)^2 \equiv \frac{1}{n} \tr
\bar{A}_i(t)
^2$.
In Fig.~\ref{fig:Rt} we plot $R(t)/R(0)$ for $N=16,32,64,128$.
It is symmetric under the time reflection $t \rightarrow -t$
as one can prove analytically even at finite $N$.
For each $N$, we have chosen
the Lagrange multiplier $\lambda$, which determines the scale of $t$,
so that $R(t)$ scales around $t=0$.
We have fixed $\lambda=1$ for $N=16$ without loss of generality.
Then we obtain $\lambda=0.92,0.72,0.59$ for $N=32,64,128$, respectively.
As we increase $N$,
the scaling region extends to larger $|t|$.
The solid line is a fit to the Gaussian function.
Thus we find that the time-evolution of the space is compatible with
the expanding behavior observed in the Monte Carlo simulation \cite{Kim:2011cr}.

\begin{figure}[htb]
\begin{center}
\includegraphics[height=6cm]{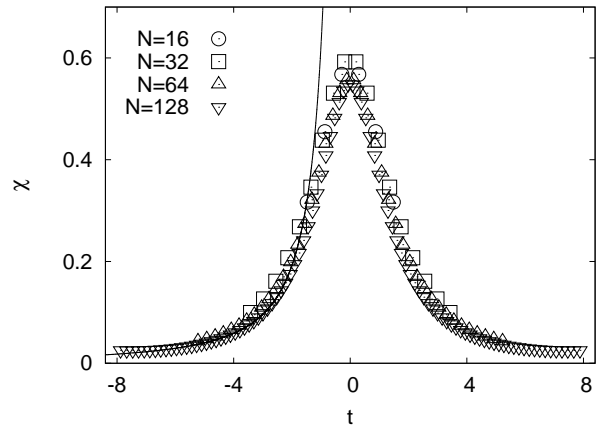}
\end{center}
\caption{
The dimensionless parameter $\chi(t)$ representing
the space-time noncommutativity is plotted against $|t|$
for $N=16,32,64,128$.
We have used the same set of values of $n$ and $\lambda$ for each $N$
as in the previous figure.
The data show nice scaling in $N$, and the solid line
represents $\chi = 0.629 \, |t|^{-1.708}$ obtained by fitting the
$N=128$ data within $-6<t<-2$ to a power law.
}
\label{fig:nc}
\end{figure}

Let us next turn our attention to the space-time noncommutativity.
We define the dimensionless parameter
\beq
\chi(t) = \frac{ - \frac{1}{n}\tr [\bar{A}_0(t) , \bar{A}_1(t)]^2  }
{ \frac{1}{n}\tr \bar{A}_0(t) ^2 \cdot  \frac{1}{n}\tr \bar{A}_1(t)^2 }
\eeq
as an estimate on the space-time noncommutativity \cite{endnote-stnc}.
In Fig.~\ref{fig:nc} we plot $\chi(t)$ for $N=16,32,64,128$.
We find that it is of O(1) at $t=0$ and
decreases as $\sim |t|^{-p}$ at large $|t|$, where we obtain
$p=1.708(3)$ by fitting the $N=128$ data within $-6<t<-2$.
Therefore, the space-time noncommutativity is significant only around
$t\sim 0$, and it becomes smaller as we go back in time.

In both Figs.~\ref{fig:Rt} and \ref{fig:nc},
we consider it very important to study the larger $|t|$ region
by increasing $N$.
In fact some preliminary results suggest certain deviation from the
Gaussian behavior of $R(t)$ and the power-law behavior of $\chi(t)$.
We hope to report on it in future publications.



\paragraph*{Summary and discussions.---} 

We have studied the classical equations of motion
in the Lorentzian
matrix model for type IIB superstring theory.
Restricting ourselves to the class of solutions
that are written in terms of Lie algebras with
four generators, we find that the solution with
SO(3) symmetry is essentially unique.
The space-time structure extracted from the solution
exhibits the time dependence, which is compatible
with the behavior observed in our previous Monte Carlo results.
The space-time noncommutativity becomes significant only towards
the end of the expansion,
whereas the space-space noncommutativity is zero.
These results suggest the appearance of an expanding
(3+1)-dimensional (almost commutative) space-time
from the Lorentzian matrix model.

We speculate that
the noncommutativity of space which plays a crucial role in
making three directions expand at earlier time, somehow
disappears at some point due to some dynamical reason.
For instance, let us use the model obtained after integrating out
the scale factor \cite{Kim:2011cr}. In that model we have a constraint
that requires the quantity (\ref{bosonic-action}) to vanish.
If the expansion with large space-space noncommutativity in the early time
continues for too long a period, the second term of (\ref{bosonic-action})
will be too large to satisfy the constraint (\ref{bosonic-action})$=0$.
Such an effect may lead
to an end of the noncommutative expansion.
One might speculate that this corresponds to the end of ``inflation''.

Our classical solution is symmetric under time reflection,
and the size of the space becomes maximum at $t=0$,
after which it has a contracting behavior.
At $t=0$, the dimensionless space-time noncommutativity becomes maximum, too,
and it is of the order one.
Hence the physics there will be quite exotic.
This may be taken as a prediction
on the fate of our universe from the Lorentzian matrix model
given that our classical solution is valid around $t=0$.

Obviously one can generalize our solution
to SO($d$) symmetric ones with $1 \le d \le 9$.
The time evolution of the size of the space and that of the
space-time noncommutativity are essentially the same as in the SO(3) case.
Let us recall here that
the space-space noncommutativity seems to play a crucial
role in the SSB of SO(9) rotational symmetry
as the mechanism proposed
in our previous work \cite{Kim:2011cr} suggests.
It is therefore not surprising that the dimensionality of space
is not fixed by classical solutions without space-space
noncommutativity.

Some comments on related works are in order.
In Ref.~\cite{Hanada:2005vr} the Einstein equation
was derived from the classical equations of motion of
the matrix model. In this derivation, however,
the matrices $A_\mu$ were interpreted as the covariant derivative
on a curved space.
It would be interesting to clarify
the relationship to our work.

Ref.~\cite{Steinacker:2011wb} reports on interesting solutions
to the classical equations of motions (\ref{cl-eq}) with
$\lambda = \tilde{\lambda}=0$.
They represent a flat Minkowski space with extra dimensions
described by fuzzy spheres. An interesting feature of these
solutions is that there exists noncommutativity between
the four extended directions and the extra dimensions.
This is crucial for
realizing a nontrivial structure
in the extra dimensions even without the Myers-like term.
Here we emphasize that the nonzero $\lambda$, which is introduced
in our work, is crucial for the expanding behavior.
Let us recall that $\lambda$ is the Lagrange multiplier
corresponding to the infrared cutoff, which turns out to be needed
according to our previous Monte Carlo studies.

In Ref.~\cite{Freedman:2004xg}
the Matrix theory \cite{BFSS} has been applied to cosmology.
A classical solution with three expanding (commutative) directions
and six oscillating (noncommutative) directions was discussed.
(The number of expanding directions does not have to be three.)
In order to have such a solution,
the authors introduced a SO(9) symmetric tachyonic mass term,
which was interpreted as the cosmological term.
The relationship to our solution is not clear, though,
since the time is treated in a different way.
The idea to use the matrices to avoid the big-bang singularity
is also pursued in Refs.~\cite{mst}.

It is tempting to imagine that the rapid growth of $R(t)$ observed
in the present solution has something to do with the accelerating
expansion confirmed by recent cosmological observations.
The power-law expansion at earlier time
may be understood by considering the quantum
corrections around the classical solution.
It is also expected that the gauge interactions
and the matter content in the (3+1)-dimensional space are determined
by the structure in the extra dimensions \cite{Aoki:2010gv,Chatzistavrakidis:2011gs}
analogously to the case of intersecting D-brane models.
We hope the present model provides a new perspective on
particle physics beyond the standard model
as well as on cosmological models for inflation, modified gravity etc..



\paragraph*{Acknowledgments.---}
We thank H.~Aoki, S.~Iso,
H.~Kawai, Y.~Kitazawa, Y.~Sekino and H.~Steinacker
for discussions.
S.-W.K.\ is supported by Grant-in-Aid
for Scientific Research
from the Ministry of
Education, Culture, Sports, Science and Technology in Japan (No. 20105002).
J.N.\ and A.T.\ is supported in part by Grant-in-Aid
for Scientific Research
(No.\ 19340066, 19540294, 20540286 and 23244057)
from JSPS.





\begin{thebibliography}{99}
\bibitem{endnote-euclid}
The mechanism of the spontaneous breaking
in the Euclidean model \cite{NV,Anagnostopoulos:2001yb}
is quite different from that in the Lorentzian model.
In the former, the Pfaffian that appears from the fermion integral
is complex, and
it becomes real positive
for collapsed configurations.
\end{thebibliography}
\end{document}